\documentclass[twocolumn,aps,pra,superscriptaddress,showpacs,tightenlines]{revtex4-1}
\usepackage{amsmath}
\usepackage{amsfonts}
\usepackage{graphicx}
\usepackage{epsfig}
\usepackage{color}
\usepackage[colorlinks,citecolor=blue]{hyperref}

\begin{document}
\title{Manipulating counter-rotating interactions in the quantum Rabi model \\via qubit frequency modulation}

\author{Jin-Feng Huang}
\email{jfhuang@hunnu.edu.cn}
\affiliation{Key Laboratory of Low-Dimensional Quantum Structures and Quantum Control of Ministry of Education, Department of Physics and Synergetic Innovation Center for Quantum Effects
and Applications, Hunan Normal University, Changsha 410081, China}
\affiliation{School of Natural Sciences, University of California, Merced, California 95343, USA}

\author{Jie-Qiao Liao}
\affiliation{School of Natural Sciences, University of California, Merced, California 95343, USA}

\author{Lin Tian}
\affiliation{School of Natural Sciences, University of California, Merced, California 95343, USA}

\author{Le-Man Kuang}
\affiliation{Key Laboratory of Low-Dimensional Quantum Structures and Quantum Control of Ministry of Education, Department of Physics and Synergetic Innovation Center for Quantum Effects
and Applications, Hunan Normal University, Changsha 410081, China}

\begin{abstract}
We propose a practical approach to manipulate the counter-rotating (CR) interactions in the quantum Rabi model by introducing a sinusoidal modulation to the transition frequency of the quantum two-level system in this model. By choosing appropriate modulation frequency and amplitude, enhancement and suppression of the CR interactions can be achieved in the Jaynes-Cummings regime (including both weak- and strong-coupling cases) as well as the ultrastrong-coupling regime. In particular, we calculate the output photon emission of the cavity vacuum state under enhanced CR terms. Our results show that continuous and steady photon emission from the cavity vacuum can be observed in the Jaynes-Cummings regime as a consequence of this enhancement. Our approach can be realized in superconducting quantum circuits.
\end{abstract}


\date{\today}
\maketitle

\section{Introduction}

The quantum Rabi model~\cite{Rabi1936,Rabi1937}, one of the fundamental models in quantum optics, describes the light-matter interaction between a quantum two-level system (atom, qubit) and a bosonic field (optical mode,  microwave mode). Based on the coupling strength between the two-level system and the quantum field, the quantum Rabi model possesses two important parameter regimes. For an interaction strength much smaller than the frequencies of the atom and the field (which are near resonance), the quantum Rabi Hamiltonian can be reduced to the Jaynes-Cummings (JC) Hamiltonian~\cite{JC1963,Knight1993JMO} under the rotating-wave approximation (RWA), i.e., the counter-rotating (CR) terms in the interaction can be omitted, and the model can be solved analytically with simple functions. This is the so-called JC regime, which can be further divided into weak- and strong-coupling cases~\cite{JCregimenote}. Cavity-QED systems formed by natural atoms coupled to optical cavities are well within this regime, which makes the JC model one of the most significant models in quantum optics. The other interesting regime is the ultrastrong-coupling regime, where the light-matter coupling strength reaches a considerable fraction of the atom and the cavity frequencies. In this regime, the RWA is no longer valid, and the CR interactions strongly modify the eigenenergies of the JC model~\cite{Irish2007PRL,Zheng2010EPJD,Braak}. To date, the ultrastrong-coupling regime has been demonstrated in superconducting circuits ~\cite{Blais2009PRA,Gross2010NatPhy,Mooij2010PRL} and semiconductor microcavities~\cite{Beltram2009PRB,Huber2009Nat,Sirtori2010PRL}, with coupling strengths exceeding $10\%$ of the qubit or the cavity frequency. Moreover, the ultrastrong-coupling regime can be simulated with various cavity-QED~\cite{Parkins2013PRA} and circuit-QED~\cite{SolanoPRX2012} systems. Recently, even the deep-strong-coupling regime has been reported, with a coupling strength comparable or larger than the qubit or the cavity frequency~\cite{Semba2016}.

In the past few years, enormous effort has been devoted towards the study of the ultrastrong-coupling regime in the quantum Rabi model. It was found that the CR terms can induce novel quantum phenomena: such as asymmetry of the vacuum Rabi-splitting~\cite{Cao2011NJP}, virtual-photon-induced vacuum Rabi oscillation~\cite{Law2013PRA}, superradiance transition~\cite{Ashhab2013PRA}, non-classical photon statistics~\cite{Hartmann2012PRL,Hartmann2013PRL}, spontaneous release of virtual photons~\cite{Savasta2013PRL,Huang2014}, and multi-photon sideband transitions~\cite{You2016}. Moreover, interesting quantum dynamical phenomena have been reported, such as collapse and revival of quantum states~\cite{Solano2010PRL}, quantum Zeno and anti-Zeno effects~\cite{Cao2010PRA,Ai2010PRA}, single-photon scattering~\cite{Wang2012PRA,Moreno2014PRL}, collective spontaneous  emission in multi-atom systems~\cite{Scully2009PRL,Rohlsberger2010Sci,Li2013PRA}, and multiphoton quantum Rabi oscillations~\cite{Garziano2015}. Owing to the versatile effects generated by the CR interactions, it becomes an interesting topic to develop techniques to control the CR interactions in the quantum Rabi model~\cite{Huang2015}. Manipulating the CR interactions can help us observe virtual photon effects in weak- or strong-coupling systems; whereas in ultrastrong-coupling regime, these techniques can help us suppress virtual photon effects. For example, Liberato~\textit{et al.} proposed a method to enhance virtual photon processes, which requires the coupling strength to oscillate at twice the frequency of the cavity mode~\cite{Liberato2009PRA}. Huang and Law proposed a scheme to control the CR interactions using a sequence of phase kicks~\cite{Huang2015}.

In this paper, we propose a practical approach to manipulate the light-matter coupling in the quantum Rabi model by introducing a monochromatic modulation of the transition frequency of the quantum two-level system. This frequency modulation induces a series of sidebands in the spectrum of the quantum two-level system, and affects the detunings between the two-level system and the cavity mode. By engineering appropriate modulation frequency and amplitude, the detunings and the effective coupling strengths between the two-level system and the cavity can be controlled to make desired transitions on resonance and unwanted transitions far off-resonance. With this approach, we can enhance the CR interactions in the weak- or strong-coupling regime, and suppress the CR interactions in the ultrastrong-coupling regime. We show that in the JC regime, the enhanced CR interactions can induce stronger photon emission from cavity vacuum state than that in previous studies~\cite{Savasta2013PRL,Huang2014}.

The paper is organized as follows. In Sec.~\ref{system}, we describe the quantum Rabi model in the presence of a frequency modulation of the quantum two-level system. In Secs.~\ref{enhancement} and~\ref{suppression}, we present approaches to enhance and suppress the CR interactions in the JC regime and the ultrastrong-coupling regime, respectively. We also study the photon emission from cavity vacuum in the enhancement case by calculating the output photon flux rate. Discussions and conclusions are given in Sec.~\ref{conclusion}.

\begin{figure}[tbp]
\center
\includegraphics[bb=30 550 250 774, width=0.47 \textwidth]{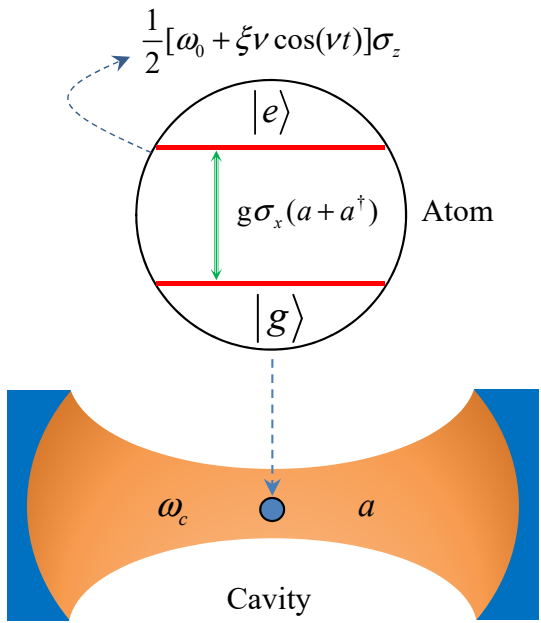}
\caption{(Color online) Schematic of the quantum Rabi model. A quantum two-level system with an energy separation $\omega_{0}$ between its excited state $|e\rangle$ and ground state $|g\rangle$ is coupled to a cavity mode with resonance frequency $\omega_{c}$. The coupling between the two-level system and the cavity has the form $g\sigma_{x}(a+a^{\dag})$. The energy separation of the two-level system is modulated with a term $(1/2)\xi\nu\cos(\nu t)\sigma_{z}$.}
\label{Fig1}
\end{figure}

\section{The system\label{system}}

We consider the quantum Rabi model, which contains a quantum two-level system interacting with a cavity mode (Fig.~\ref{Fig1}). The Hamiltonian of the Rabi model can be written as $(\hbar=1)$~\cite{Rabi1936,Rabi1937}
\begin{eqnarray}
H_{R}=H_{\textrm{JC}}+H_{\textrm{CR}},
\end{eqnarray}
where
\begin{equation}
H_{\textrm{JC}}=\omega_{c}a^{\dagger}a+\frac{\omega_{0}}{2}\sigma_{z}+g(\sigma_{+}a+a^{\dagger}\sigma_{-}),
\end{equation}
is the usual JC Hamiltonian, and
\begin{equation}
H_{\textrm{CR}}=g(\sigma_{-}a+a^{\dagger}\sigma_{+}),
\end{equation}
includes the CR terms. Here, $a$ $(a^{\dagger})$ is the annihilation (creation) operator of the cavity field with frequency $\omega_{c}$. The quantum two-level system is described by the Pauli operators $\sigma_{z}=|e\rangle\langle e|-|g\rangle\langle g|$ and $\sigma_{+}=\sigma_{-}^{\dagger}=|e\rangle\langle g|$, where $\vert e\rangle$ and $\vert g\rangle$ are the excited state and the ground state, respectively, with an energy separation $\omega_{0}$. The $g$ terms in $H_{R}$ describe the interactions between the two-level system and the cavity field. It should be pointed out that our model is general, and can be implemented with various experimental setups.

To manipulate the CR terms in the light-matter interactions, we introduce a sinusoidal modulation to the energy separation of the two-level system. The modulation Hamiltonian is given by
\begin{equation}
H_{M}(t)=\frac{1}{2}\xi\nu\cos(\nu t)\sigma_{z},
\end{equation}
where $\nu$ and $\xi$ are the modulation frequency and normalized modulation amplitude, respectively. We note that frequency modulation has been recently studied in various tasks in quantum optics~\cite{Shevchenko2010PR,Silveri2015,Zhou2009PRA,Beaudoin2012PRA,Strand2013PRB,Li2013NC,Liao2015PRA,Liao2016PRA}.
The total Hamiltonian of this system under the frequency modulation is then $H(t)=H_{R}+H_{M}(t)$, which can be divided as the following:
\begin{equation}
H(t)=H_{0}(t)+H_{I},\label{Hamiltot}
\end{equation}
with
\begin{subequations}
\begin{align}
H_{0}(t)&=\omega_{c}a^{\dagger}a+\frac{1}{2}[\omega_{0}+\xi\nu\cos(\nu t)]\sigma_{z},\\
H_{I}&=g\sigma_{x}(a+a^{\dagger}).
\end{align}
\end{subequations}
Here $H_{0}(t)$ is the time-dependent non-interacting Hamiltonian of the two-level system and the cavity mode, and $H_{I}$ describes the light-matter coupling.

To study the impact of the frequency modulation on the dynamics of this system, we perform the following transformation on the system
\begin{eqnarray}
V(t)&=&\mathcal{T}\exp\left[-i\int^{t}_{0}H_{0}(\tau)d\tau\right]\nonumber\\
&=&\exp\left\{-i\omega_{c}ta^{\dagger}a-i\frac{1}{2}[\omega_{0}t+\xi\sin(\nu t)]\sigma_{z}\right\},
\end{eqnarray}
where $\mathcal{T}$ denotes the time-ordering operator. In the rotating frame defined by $V(t)$, the transformed Hamiltonian becomes
\begin{eqnarray}
\tilde{H}(t)&=& V^{\dagger}(t)H(t)V(t)-iV^{\dagger}(t)\dot{V}(t)\nonumber \\
&=&g\left(\sigma_{+}e^{i[\omega_{0}t+\xi\sin(\nu t)]}+\sigma_{-}e^{-i[\omega_{0}t+\xi\sin(\nu t)]}\right)\nonumber\\
&&\times(ae^{-i\omega_{c}t}+a^{\dagger}e^{i\omega_{c}t})\nonumber\\
&=&\sum_{n=-\infty}^{\infty}gJ_{n}(\xi)[\sigma_{+}ae^{i(\delta+n\nu)t}+\textrm{H.c.}]\nonumber \\
&&+\sum_{m=-\infty}^{\infty}gJ_{m}(\xi)[\sigma_{+}a^{\dagger}e^{i\Delta_{m}t}+\textrm{H.c.}],\label{eq:H1}
\end{eqnarray}
where $\delta=\omega_{0}-\omega_{c}$ is the detuning between the unmodulated two-level system and the cavity, $J_{n}(\xi)$ is the $n$th Bessel function of the first kind, and $\Delta_{m}=\omega_{0}+\omega_{c}+m\nu$. In the derivation of $\tilde{H}(t)$, we have used the Jacobi-Anger expansion,
\begin{equation}
e^{i\xi\sin(\nu t)}=\sum_{n=-\infty}^{\infty}J_{n}(\xi)e^{in\nu t}.\label{eq:Jacobi}
\end{equation}
The rotating and CR terms in Hamiltonian $\tilde{H}(t)$ can be tailored by choosing appropriate modulation parameters $\xi$ and $\nu$. In particular, the normalized coupling strengths $gJ_{n(m)}(\xi)$ can be changed in a large range by tuning $\xi$, and the detunings for different sidebands $\delta+n\nu$ and $\Delta_{m}$ can be controlled by adjusting the modulation frequency $\nu$ and the sideband parameters $n$ ($m$). This controllability enables us to enhance the CR terms in the weak- or strong-coupling regime and suppress the CR terms in the ultrastrong-coupling regime.

\section{\label{enhancement}Enhancement of the CR terms}
In this section, we study the enhancement of the CR terms in the JC regime, where, roughly speaking, $g< 0.1\omega_{0}$, $0.1\omega_{c}$. In this regime and under the near-resonance condition $|\delta|\ll\omega_{0}$, $\omega_{c}$, the CR interactions $H_{\textrm{CR}}$ in the Rabi Hamiltonian $H_{R}$ can be safely omitted by applying the RWA, and $H_{R}$ is then reduced to the JC Hamiltonian $H_{\textrm{JC}}$. In the following, we will show how to enhance the CR interactions by introducing frequency modulation to the two-level system.

\subsection{Effective Hamiltonian and dynamics}

We first analyze the modulated interactions in Eq.~(\ref{eq:H1}). For the rotating terms (the fourth line) in Eq.~(\ref{eq:H1}), the detunings of the sidebands ($\delta+n\nu$) are separated from each other by $(n-n^{\prime}) \nu$ with $n,\,n^{\prime}$ being integers. Under the condition $\nu\gg g>g|J_{n}(\xi)|$, all the rotating terms other than the $0$th-order sideband with the effective coupling strength $gJ_{0}(\xi)$ can be discarded using the RWA. To enhance the CR interactions (the fifth line) in Eq.~(\ref{eq:H1}), we choose a modulation frequency $\nu$ such that there exists a CR sideband $m=m_{0}$ that satisfies the near-resonance condition, i.e., $|\Delta_{m_{0}}|< |gJ_{m_{0}}(\xi)|$. For this sideband, the CR term has a comparable effect as that of the resonant rotating term. The detunings of all other CR sidebands are $\Delta_{m_{0}+s}=\Delta_{m_{0}}+s\nu$, with $s$ being a nonzero integer. Under the conditions
\begin{eqnarray}
 \nu\gg  |\Delta_{m_{0}}|,\hspace{0.5 cm} \nu\gg g>g|J_{m}(\xi)|,\label{appxcond1}
\end{eqnarray}
all the other CR terms can be discarded with the RWA. The Hamiltonian $\tilde{H}(t)$ can hence be approximated as
\begin{eqnarray}
\tilde{H}_{1}(t)\approx (g_{r}\sigma_{+}ae^{i\delta t}+g_{c}\sigma_{+}a^{\dagger}e^{i\Delta_{m_{0}}t})+\textrm{H.c.},\label{eq:H1app}
\end{eqnarray}
where $g_{r}=gJ_{0}(\xi)$ and $g_{c}=gJ_{m_{0}}(\xi)$ are the normalized coupling strengths for the resonant rotating and CR terms, respectively. The Hamiltonian~(\ref{eq:H1app}) describes an effective isotropic Rabi model with $g_{r}= g_{c} $ or an effective anisotropic Rabi model with $g_{r}\neq g_{c} $~\cite{Xie2014PRX} with a cavity frequency $\tilde{\omega}_{c}=(\Delta_{m_{0}}-\delta)/2$ and a frequency $\tilde{\omega}_{0}=(\Delta_{m_{0}}+\delta)/2$ for the quantum two-level system. We can choose appropriate $\Delta_{m_{0}}$ and $\delta$ such that the effective Rabi Hamiltonian $\tilde{H}_{1}(t)$ enters the ultrastrong-coupling regime with
$|g_{c}|$ reaches $0.1$ of the frequencies $\tilde{\omega}_{c}$ and $\tilde{\omega}_{0}$.
In particular, a resonant CR interaction can be obtained at $\Delta_{m_{0}}=0$, which requires that
\begin{equation}
\nu=-\frac{\omega_{0}+\omega_{c}}{m_{0}}, \label{rescond1}
\end{equation}
at selected value of $m_{0}$ with $m_{0}$ being a negative integer. This indicates that the modulated CR interactions can be enhanced even though they are negligibly weak in the original representation.

\begin{figure}
\includegraphics[bb =0 3 408 529, width=0.47 \textwidth]{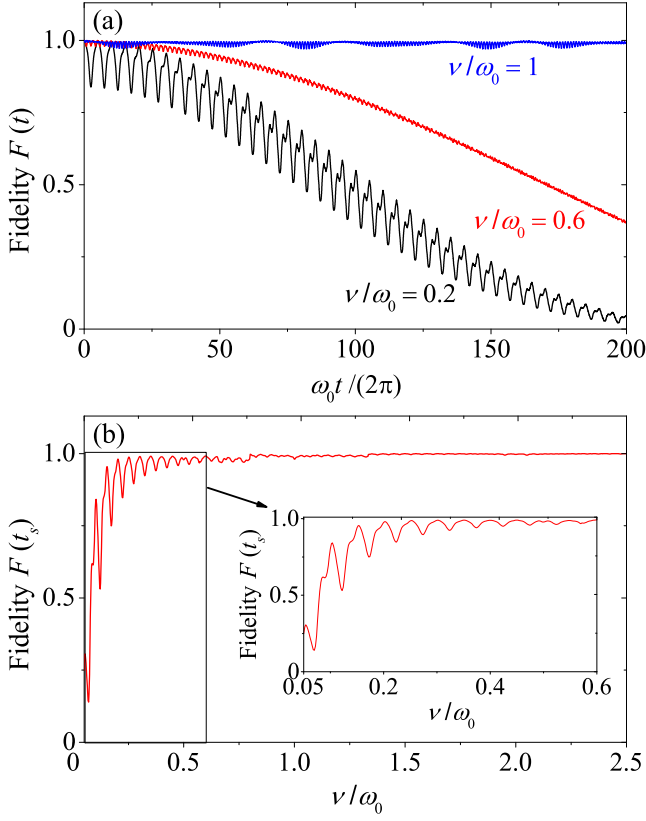}
\caption{(Color online) (a) Fidelity $F(t)$ defined in Eq.~(\ref{fidelity}) versus the time for various values of the modulation frequency: $\nu=0.2\omega_{0}=4g$ (black solid curve), $\nu=0.6\omega_{0}=12g$ (red solid curve), and $\nu=\omega_{0}=20g$ (blue solid curve). (b) Fidelity $F(t_{s})$ at time $t_{s}=2\pi/g$ as a function of $\nu$. Other parameters are $\xi=2.40483$, $\omega_{c}=\omega_{0}$, and $g=0.05\omega_{0}$. The initial state of the system is $(1/\sqrt{2})(|g\rangle+|e\rangle)|\alpha\rangle$, where $|\alpha\rangle$ is a coherent state with $\alpha=0.1$.}
\label{Fig2}
\end{figure}
The validity of the RWA in deriving the effective Hamiltonian~(\ref{eq:H1app}) can be checked by studying the fidelity
\begin{equation}
F(t)=|\langle\phi(t)|\psi(t)\rangle|^{2}\label{fidelity}
\end{equation}
between the state $|\phi(t)\rangle$ obtained by solving the Schr\"{o}dinger equations with the exact Hamiltonian (\ref{eq:H1}) and the state $|\psi(t)\rangle$, obtained from the effective Hamiltonian (\ref{eq:H1app}). Here we choose an initial state $|\phi(0)\rangle=|\psi(0)\rangle=(1/\sqrt{2})(|g\rangle+|e\rangle)|\alpha\rangle$, where $|\alpha\rangle$ is a coherent state of the cavity field, to calculate the fidelity.

In Fig.~\ref{Fig2}(a), we plot the fidelity $F(t)$ as a function of time for several values of the modulation frequency $\nu$ starting from the initial state  $|\phi(0)\rangle$. Here, for a given $\nu$, the sideband integer $m_{0}$ is chosen such that the corresponding CR term is the most resonant term, i.e., $m_{0}=\texttt{Round}[-(\omega_{0}+\omega_{c})/\nu]$, where $\texttt{Round}[x]$ is a function for getting the nearest integer of $x$. Figure~\ref{Fig2}(a) shows that the fidelity experiences oscillations with time. When $\nu$ is much smaller than $\omega_{0}$, such as $\nu=0.2\omega_{0}=4g$, the envelope of the fidelity decreases with time accompanied by periodic revival in a long range of time. For larger $\nu$, better fidelity can be obtained, which is in accordance with conditions~(\ref{appxcond1}). To find out how the fidelity depends on the frequency $\nu$, in Fig.~\ref{Fig2}(b), we plot the fidelity $F(t_{s})$ at time $t_{s}=2\pi/g$ as a function of $\nu$. We see that the fidelity at $t_{s}$ experiences a fast oscillation for $\nu/\omega_{0}<0.8$. However, the envelope of the fidelity increases gradually with the modulation frequency $\nu$. For $\nu>\omega_{0}=20g$, a high fidelity [$F(t_{s})\approx1$] can be obtained.

\begin{figure}[tbp]
\center
\includegraphics[bb=4 1 266 284, width=0.47 \textwidth]{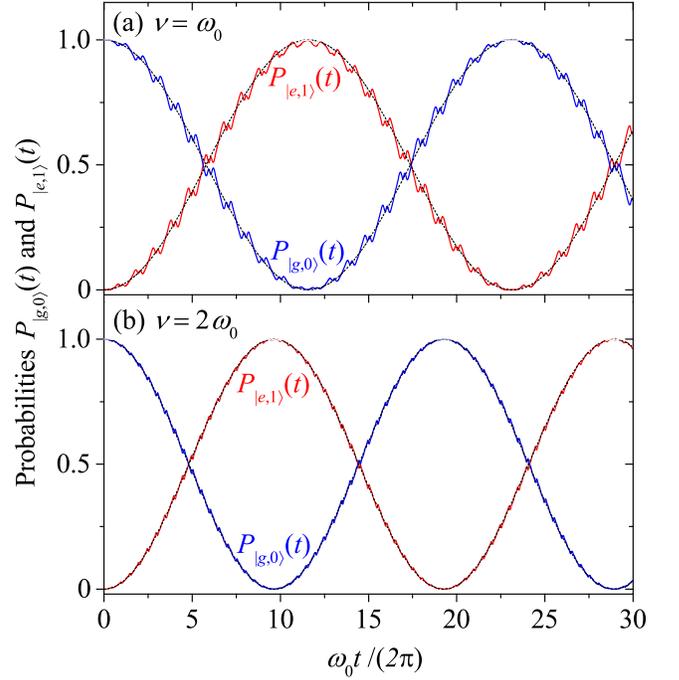}
\caption{(Color online) Time dependence of the probabilities $P_{|g,0\rangle}(t)$ (blue solid curves) and $P_{|e,1\rangle}(t)$ (red solid curves) obtained with the exact Hamiltonian $\tilde{H}(t)$ when the modulation frequency $\nu$ satisfies sideband resonance condition at two different values: (a) $\nu=\omega_{0}$ ($m_{0}=-2$), (b) $\nu=2\omega_{0}$ ($m_{0}=-1$). The black short-dashed curves are the analytical results in Eqs.~(\ref{RbiaJCana}) obtained with the effective Hamiltonian $\tilde{H}_{\textrm{aJC}}(t)$. Other parameters are $J_{0}(\xi)=0$ ($\xi=2.40483$), $g=0.05\omega_{0}$, and
$\omega_{c}=\omega_{0}$. The initial state of the system is $|g,0\rangle$.}
\label{Fig3}
\end{figure}
The above technique can be used to produce pure CR interactions. By choosing proper values of $\xi$, the rotating terms in Eq.~(\ref{eq:H1app}) disappear. For example, at $\xi=2.40483$, $J_{0}(\xi)=0$, and the Hamiltonian~(\ref{eq:H1app}) becomes
\begin{equation}
\tilde{H}_{\textrm{aJC}}(t)\equiv g_{c}(\sigma_{+}a^{\dagger}e^{i\Delta_{m_{0}} t}+a\sigma_{-}e^{-i\Delta_{m_{0}} t}),\label{eq:HaJC}
\end{equation}
only containing the CR terms. An essential feature associated with this Hamiltonian is the Rabi oscillation between the states $|g,0\rangle$ and $|e,1\rangle$, where $|n=0,1\rangle$ are number states of the cavity field. Note that multiphoton Rabi oscillations in ultrastrongly-coupled cavity-QED systems have recently been considered~\cite{Garziano2015}. With $\tilde{H}_{\textrm{aJC}}(t)$ and the initial state $|g,0\rangle$, the probabilities of the system in states $|g,0\rangle$ and $|e,1\rangle$ can be obtained as
\begin{subequations}
\label{RbiaJCana}
\begin{align}
P_{|g,0\rangle}(t)&=\frac{4g^{2}_{c}}{4g^{2}_{c}+\Delta^{2}_{m_{0}}}\cos^{2}\left(\frac{1}{2}\sqrt{4g^{2}_{c}+\Delta^{2}_{m_{0}}}t\right),\\
P_{|e,1\rangle}(t)&=\frac{4g^{2}_{c}}{4g^{2}_{c}+\Delta^{2}_{m_{0}}}\sin^{2}\left(\frac{1}{2}\sqrt{4g^{2}_{c}+\Delta^{2}_{m_{0}}}t\right).
\end{align}
\end{subequations}
This Rabi oscillation can be utilized to evaluate the validity of the RWA performed in this modulation scheme. To this end, we compare the exact dynamics of the system governed by the full Hamiltonian~(\ref{eq:H1}) with the analytical solution in Eq.~(\ref{RbiaJCana}).

In Fig.~\ref{Fig3}, we show the exact and approximate results at two different resonance sidebands, i.e., different values of $m_{0}$ under $\Delta_{m_{0}}=0$. Figures~\ref{Fig3}(a) and~\ref{Fig3}(b) correspond to $m_{0}=-2$ and $-1$, respectively. Figure~\ref{Fig3} shows that the approximate result agrees well with the exact dynamics when the conditions~(\ref{appxcond1}) are satisfied. The system demonstrates a clear Rabi oscillation (with period $\pi/|g_{c}|$) between the states $|g,0\rangle$ and $|e,1\rangle$, and the leakage of the system out of this subspace (spanned by $|g,0\rangle$ and $|e,1\rangle$) is negligible. These features are evidences of the validity of the RWA, and clearly show the enhancement of CR interactions under well-designed frequency modulation. In the absence of modulation (when $H_{M}=0$), the system will stay in the state $|g,0\rangle$, which is the ground state of the Hamiltonian $H_{R}$ under the RWA.

\begin{figure}[tbp]
\center
\includegraphics[bb = 14 5 341 266.5, width=0.47 \textwidth]{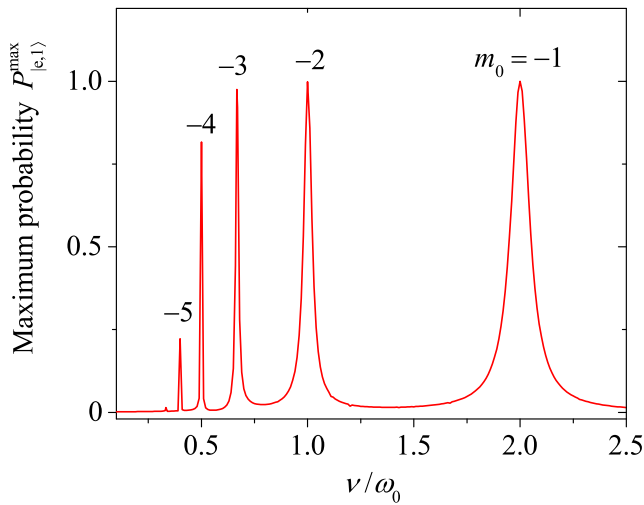}
\caption{(Color online) Maximum probability $P_{|e,1\rangle}^{\textrm{max}}$
of state $|e,1\rangle $ versus the modulation frequency $\nu$. Other parameters are the same as those in Fig.~\ref{Fig3}.}
\label{Fig4}
\end{figure}

The above sideband-resonance effect can be illustrated more clearly by studying the dependence of the maximum probability $P_{|e,1\rangle}^{\textrm{max}}$ on the modulation frequency $\nu$. Let's consider the fifth line of Eq.~(\ref{eq:H1}).
As we sweep the modulation frequency $\nu$, there exist a sequence of resonance windows which correspond to different values of $m_{0}$ when the resonance condition~(\ref{rescond1}) ($\Delta_{m_{0}}=0$) is satisfied. At resonance, the effective Hamiltonian becomes (\ref{eq:HaJC}) by applying the RWA with the conditions~(\ref{appxcond1}) and $J_{0}(\xi)=0$. In this case, the maximum probability of state $|e,1\rangle$ in the Rabi oscillation can be obtained. In Fig.~\ref{Fig4}, we plot the maximum probability $P_{|e,1\rangle}^{\textrm{max}}$ as a function of $\nu$ by solving the full Hamiltonian~(\ref{eq:H1}) with the same parameters as those used in Fig.~\ref{Fig3}. We see from Fig.~\ref{Fig4} that there are resonance peaks at locations
predicted by Eq.~(\ref{rescond1}).

The linewidths of the resonance peaks in Fig.~\ref{Fig4} can be obtained by analyzing the resonance condition. At exact resonance with $\Delta_{m_{0}}=0$, the maximum value of the probability is almost one: $P^{\textrm{max}}_{|e,1\rangle}\approx1$. According to Eq.~(\ref{RbiaJCana}), with the increase of the detuning $|\Delta_{m_{0}}|$, the value of $P^{\textrm{max}}_{|e,1\rangle}$ decreases approximately by the relation $4g^{2}_{c}/(4g^{2}_{c}+\Delta^{2}_{m_{0}})$, which is a Lorentzian function of $|\Delta_{m_{0}}|$. Therefore the linewidths of these peaks are determined by the full width at half maximum, i.e., $|\Delta_{m_{0}}/(2g_{c})|\approx 1$. A rough estimation gives the linewidth for the peak associated with $m_{0}$ as $|2gJ_{m_{0}}(\xi)/m_{0}|$ at $\xi=2.40483$, which decreases with the increase of $|m_{0}|$, in accordance with Fig.~\ref{Fig4}. In addition, Fig.~\ref{Fig4} shows that the height of the two peaks ($m_{0}=-5$, $-4$) from the left is smaller than one. This is because there exist some rotating terms which will affect the population significantly. For example, when $m_{0}=-4$, the rotating and CR terms at $m=-2$ have a considerable effect on the dynamics. It is worth noting that the present mechanism of enhancing the CR interactions works well for other values of detuning $\delta$. This is because the resonance condition $\Delta_{m_{0}}=0$ is independent of the detuning $\delta$.

\subsection{Output photon flux}

The above discussions are for a closed system without dissipation. In this section, we study the impact of the frequency modulation in the presence of environmental noise, namely, how the modulation affects the system transitions induced by the environment. We assume that the two-level system and the cavity are each connected to a (different) zero-temperature bath. In the absence of the modulation, the system is well described by the JC model. When the system is initially prepared in its ground state $|g,0\rangle$, the system will always stay in this state under a zero-temperature bath and there is no output photon flux from the cavity. When an appropriate modulation is applied, the system is described by the quantum Rabi model. Here when the system is initially prepared in the state $|g,0\rangle$, the dissipation induced by the zero-temperature baths will lead to a finite photon flux to the continuous fields in the cavity output. This is because the state $|g,0\rangle$ is not a ground state of the Rabi model; instead, it is a superposition of many eigenstates of this model. The dissipation will induce transitions from upper eigenstates to lower eigenstates, and the cavity will then emit photons.

To verify the above analyses, we simulate the cavity photon emission in the open-system case under the frequency modulation. In this case, the evolution of the system is governed by the quantum master equation~\cite{Hartmann2012PRL,Savasta2013PRL,Huang2014}
\begin{eqnarray}
\frac{d\rho(t)}{dt}&=&i[\rho(t),H(t)]\nonumber\\
&&+\sum_{s=a,c}\sum_{k>j}\sum_{j=1}^{\infty}\Gamma_{kj}^{(s)}\{D[|\varepsilon_{j}\rangle\langle\varepsilon_{k}|]\rho(t)\},\label{mastereq}
\end{eqnarray}
where $\rho(t)$ is the density matrix of the system in the Schr\"{o}dinger picture, $H(t)$ is the Hamiltonian of the system given by Eq.~(\ref{Hamiltot}), and $D[|\varepsilon_{j}\rangle\langle \varepsilon_{k}|]$ is a standard Lindblad superoperator defined by
\begin{equation}
D[O]\rho=O\rho O^{\dagger}-\frac{1}{2}O^{\dagger}O\rho-\frac{1}{2}\rho O^{\dagger}O,
\end{equation}
with $|\varepsilon_{j}\rangle\langle\varepsilon_{k}|$ being state transition operators among the eigenstates of the Rabi Hamiltonian: $H_{R}|\varepsilon_{n}\rangle=\varepsilon_{n}|\varepsilon_{n}\rangle$ for $n=1,2,3,\cdots$. The relaxation coefficients in Eq.~(\ref{mastereq}) are given by
\begin{equation}
\Gamma_{kj}^{(s=a,c)}=\gamma_{s}|C^{(s)}_{jk}|^{2},
\end{equation}
where $\gamma_{a}$ and $\gamma_{c}$ are the decay rates of the two-level system
and the cavity field, respectively, and $C^{(s=a,c)}_{jk}$ are the matrix elements of the operators $\sigma_{x}$
and $(a+a^{\dagger})$ in the eigen-representation of the Rabi model. These matrix elements are given by
\begin{subequations}
\begin{align}
C^{(a)}_{jk}&=\langle\varepsilon_{j}|\sigma_{x}|\varepsilon_{k}\rangle,\\
C^{(c)}_{jk}&=\langle\varepsilon_{j}|(a+a^{\dagger})|\varepsilon_{k}\rangle.
\end{align}
\end{subequations}

\begin{figure}[tbp]
\center
\includegraphics[bb = 5 2 265 282, width=0.47 \textwidth]{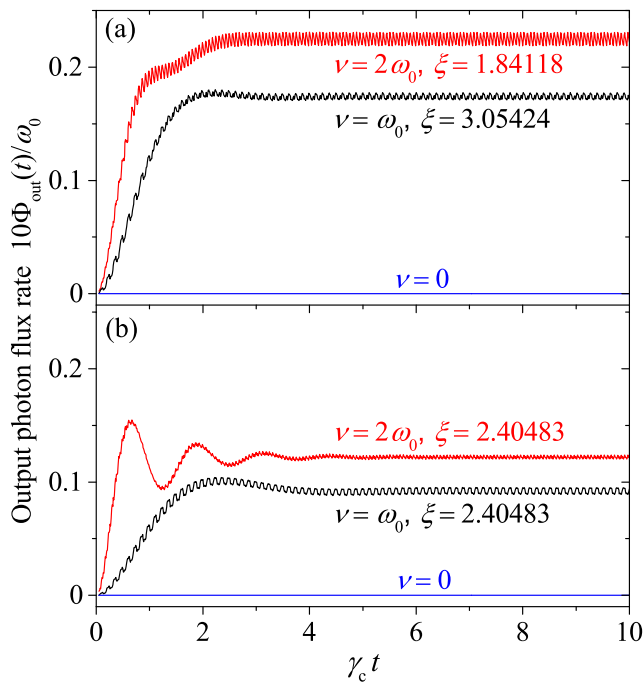}
\caption{(Color online) Time dependence of the output photon flux rate $\Phi_{\textrm{out}}(t)$ in various cases with resonant CR interactions ($\Delta_{m_{0}}=0$) but different coupling strengths $g_{r}$ and $g_{c}$. Corresponding to $\nu=\omega_{0}$ ($2\omega_{0}$), the sideband parameter of the resonant term is $m_{0}=-2$ ($-1$). (a) The parameter $\xi$ is chosen such that the coupling strength $|g_{c}|=g|J_{m_{0}}(\xi)|$ corresponding to the resonant $m_{0}$ term is maximized. (b) The parameter $\xi=2.40483$ is chosen such that the rotating terms in Eq.~(\ref{eq:H1app}) disappears [$J_{0}(\xi)=0$] and the Hamiltonian is given by Eq.~(\ref{eq:HaJC}). In these two panels, the unmodulated case $\nu=0$ is presented for comparison. Other parameters are $\delta=0$, $g=0.05\omega_{0}$, and $\gamma_{a}=\gamma_{c}=0.02\omega_{0}$. The initial state of the system is $|g,0\rangle$.}
\label{Fig5}
\end{figure}

In the open-system case, the system will transit from upper eigenstates to lower eigenstates and photons will be released from the cavity. The output photon flux can be calculated in terms of the input-output relation. In the ultrastrong-coupling regime, the input-output relation is defined by~\cite{Hartmann2012PRL,Savasta2013PRL,Huang2014}
\begin{equation}
B_{\textrm{out}}(t)=B_{\textrm{in}}(t)-i\sqrt{\gamma_{c}}X_{c}(t),
\end{equation}
where $B_{\textrm{in}}(t)$ and $B_{\textrm{out}}(t)$ are, respectively, the input and output operators, and $X_{c}(t)$ is the lowering operator in the Heisenberg picture.

In the absence of external cavity driving, the output photon flux rate can be obtained as~\cite{Savasta2013PRL,Huang2014}
\begin{equation}
\Phi_{\textrm{out}}(t)=\langle B^{\dagger}_{\textrm{out}}(t)B_{\textrm{out}}(t)\rangle=\gamma_{c}\textrm{Tr}[\rho(t)X_{c}^{\dagger}X_{c}],
\end{equation}
where
\begin{eqnarray}
X_{c}=\sum_{k>j}\sum_{j=1}^{\infty}C^{(c)}_{jk}|\varepsilon_{j}\rangle\langle \varepsilon_{k}|
\end{eqnarray}
is the lowering operator in the Schr\"{o}dinger picture.

\begin{figure}[tbp]
\center
\includegraphics[bb=14 2 340 267, width=0.47 \textwidth]{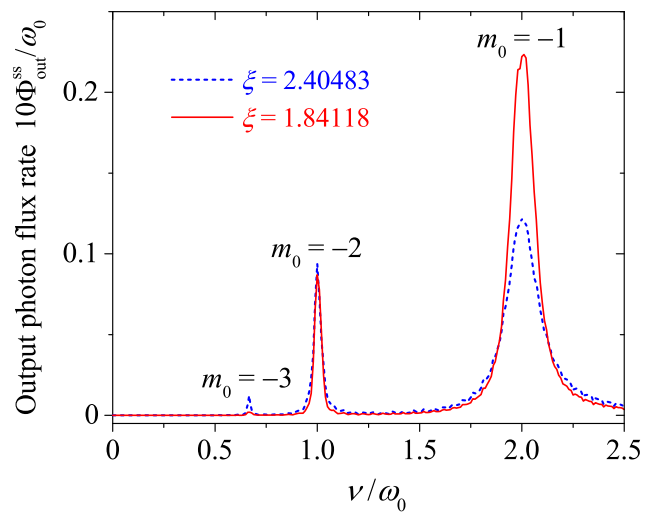}
\caption{(Color online) Steady-state output photon flux rate $\Phi_{\textrm{out}}^{\textrm{ss}}$ versus the modulation frequency $\nu$ for $\xi=1.84118$ and $\xi=2.40483$. Other parameters are $\delta=0$, $g=0.05\omega_{0}$, and $\gamma_{a}=\gamma_{c}=0.02\omega_{0}$. The initial state of the system is $|g,0\rangle$.}
\label{Fig6}
\end{figure}

In Fig.~\ref{Fig5}, we plot the time dependence of the output photon flux rate $\Phi_{\textrm{out}}(t)$ when the CR interaction in Eq.~(\ref{eq:H1app}) is at resonance. For $\nu=\omega_{0}$ ($2\omega_{0}$), the sideband parameter of the resonant term is $m_{0}=-2$ ($-1$). We also present the result of the unmodulated case $\nu=0$ for comparison. In Fig.~\ref{Fig5}(a), the parameter $\xi$ is chosen such that  the coupling strength $g_{c}$ in the corresponding resonant CR terms can be maximized. Namely, a maximum value of $J_{m_{0}}(\xi)$ is reached. When $\nu=2\omega_{0}$, the $m_{0}=-1$ term becomes resonant by $\Delta_{m_{0}}=0$, and we choose $\xi=1.84118$ to obtain a maximum value of $|J_{-1}(1.84118)|=0.581865$. Similarly, when $\nu=\omega_{0}$, the $m_{0}=-2$ term becomes resonant, and we let $\xi=3.05424$ to maximize $|J_{-2}(3.05424)|=0.486499$. From Fig.~\ref{Fig5}(a) we see that the output photon flux rate increases gradually with the increase of time. When $t\gg \pi/\gamma_{c}$, the value of $\Phi_{\textrm{out}}(t)$ approaches a stationary value with a small oscillation, which is caused by the discarded terms under the RWA. A nonzero stationary value of $\Phi_{\textrm{out}}(t)$ implies a continuous emission of real photons from the cavity vacuum in the stationary state. To illustrate the impact of the modulation, we also present the result of the unmodulated case for comparison, where no real photon is emitted. These results indicate that the enhancement of the CR interactions by the frequency modulation is a key factor in creating real photon emission in the JC regime of the light-matter coupling, contrary to that of the ultrastrong-coupling regime.

For the two cases of nonzero $\nu$ in Fig.~\ref{Fig5}(a), the resonant CR terms are maximized by choosing a proper $\xi$. However, the rotating terms also exist in Hamiltonian~(\ref{eq:H1app}). To eliminate the rotating terms and obtain pure CR interactions, in Fig.~\ref{Fig5}(b) we choose the same $\xi$ as used in Fig.~\ref{Fig2}, $\xi=2.40483$ [$J_{0}(\xi)=0$], such that the rotating terms in Hamiltonian~(\ref{eq:H1app}) can be neglected. In this case, we can observe similar features as that in Fig.~\ref{Fig5}(a) in short- and long-time limits. For intermediate time, $\Phi_{\textrm{out}}(t)$ exhibits an oscillatory behavior.

In order to illustrate the sideband resonance in the real photon emission, in Fig.~\ref{Fig6}, we plot the steady-state output photon flux rate $\Phi^{ss}_{\textrm{out}}$ as a function of the modulation frequency $\nu$ with the same parameters as used in Fig.~\ref{Fig5}. Our calculation gives resonance peaks that correspond to sidebands at different $m_{0}$. The locations of these peaks are determined by the resonance condition $\Delta_{m_{0}}=0$, similar to the probability dynamics given in Fig.~\ref{Fig4}. It is interesting to note that the magnitude of the stationary output photon flux rate depends on the coupling strength of the CR interactions. For the $m_{0}=-1$ peak ($\nu=2\omega_{0}$), the magnitude of the output photon flux rate at $\xi=1.84118$ is larger than the value at $\xi=2.40483$, because $J_{1}(\xi)$ reaches its maximum at $\xi=1.84118$.

\begin{figure}[tbp]
\center
\includegraphics[bb=5 2 405 528, width=0.47 \textwidth]{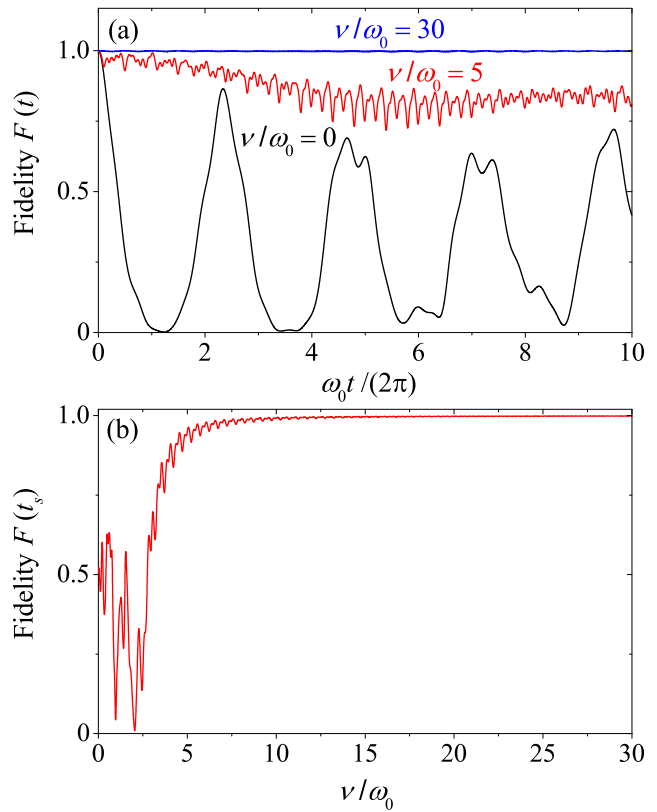}
\caption{(Color online) (a) Time dependence of the fidelity $F(t)$ at various values of modulation frequency: $\nu=0$ (black solid curve), $\nu=5\omega_{0}$ (red solid curve), and $\nu=30\omega_{0}$ (blue solid curve). (b) Fidelity $F(t_{s})$ at time $t_{s}=2\pi/g$ as a function of $\nu$. Other parameters are $\xi=2.21868$, $\omega_{c}=\omega_{0}$, and $g=0.5\omega_{0}$. The initial state of the system is $(1/\sqrt{2})(|g\rangle+|e\rangle)|\alpha\rangle$, where $|\alpha\rangle$ is a coherent state with $\alpha=0.1$.}
\label{Fig7}
\end{figure}

\section{\label{suppression}Suppression of the CR terms}

We now turn to the ultrastrong-coupling regime: $g/\omega_{0}\geq0.1$ and $g/\omega_{c}\geq0.1$, in which the CR interactions in $H_{\textrm{CR}}$
become significant. We study how to suppress the CR interactions in the ultrastrong-coupling regime by introducing frequency modulation $H_{M}(t)$ on the quantum two-level system. Our approach is to choose appropriate modulation parameters such that the CR interactions in Eq.~(\ref{eq:H1}) become far off-resonance. More explicitly, under the parameter condition
\begin{eqnarray}
\nu\gg\omega_{0}+\omega_{c}\gg g_{r}=gJ_{0}(\xi),\label{RWAcod2}
\end{eqnarray}
the transformed Hamiltonian $\tilde{H}(t)$ can be simplified as
\begin{eqnarray}
\tilde{H}_{\textrm{JC}}(t)\approx g_{r}(\sigma_{+}ae^{i\delta t}+a^{\dagger}\sigma_{-}e^{-i\delta t}).\label{eq:Vjc}
\end{eqnarray}
This Hamiltonian describes a JC model in the interaction picture with detuning $\delta$ and effective Rabi frequency $g_{r}$.

Similar to the enhancement case in the previous section, we use a fidelity to evaluate the validity of the approximate Hamiltonian~(\ref{eq:Vjc}) compared with the original Hamiltonian~(\ref{eq:H1}). The fidelity has the same form as Eq.~(\ref{fidelity}), but with the state $|\psi\rangle$ determined by the approximate Hamiltonian~(\ref{eq:Vjc}). We also choose the initial state of the system as $(1/\sqrt{2})(|g\rangle+|e\rangle)|\alpha\rangle$, with $|\alpha=0.1\rangle$ being a coherent state. In Fig.~\ref{Fig7}(a), we plot the dynamics of the fidelity $F(t)$ at several values of the modulation frequency $\nu$. We can see that a better fidelity can be obtained for a larger $\nu$, which is in accordance with Eq.~(\ref{RWAcod2}). In particular, since the coupling strengths of the rotating and the CR terms in Eq.~(\ref{eq:H1}) are normalized to $gJ_{m}(\xi)$ (much smaller than $g$), the RWA performed in obtaining Eq.~(\ref{eq:Vjc}) is valid even under a moderate value of $\nu/\omega_{0}$ [for example $\nu/\omega_{0}=5$ in Fig.~\ref{Fig7}(a)]. This is because, for a given discarded term, its contribution is determined by the ratio of the oscillating frequencies over the coupling strength. The larger this ratio, the better the approximation. To investigate how the fidelity depends on the frequency $\nu$ more clearly, in Fig.~\ref{Fig7}(b), we plot the fidelity $F(t_{s})$ at time $t_{s}=2\pi/g$ as a function of $\nu$. Figure~\ref{Fig7}(b) shows that the curve exhibits oscillations followed by a fast increase when $\nu>2\omega_{0}$ until reaching a steady value. A high fidelity of $F(t_{s})\approx1$ can be obtained for $\nu>5\omega_{0}$.

\begin{figure}[tbp]
\center
\includegraphics[bb=14 2 340 267, width=0.47 \textwidth]{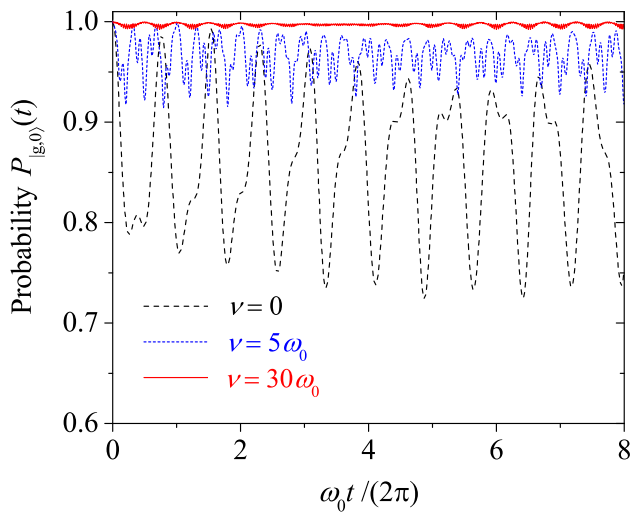}
\caption{(Color online) Time dependence of the probability $P_{|g,0\rangle}(t)$ of the system in state $|g,0\rangle$ at various values of $\nu$: $\nu=0$ (black dashed curve), $\nu=5\omega_{0}$ (blue short-dashed curve), and $\nu=30\omega_{0}$ (red solid curve). Other parameters are $\xi=2.21868$ [$J_{0}(\xi)=0.1$], $g=0.5\omega_{0}$, and $\omega_{c}=\omega_{0}$. The initial state of the system is $|g,0\rangle$.}
\label{Fig8}
\end{figure}
To illustrate the suppression of the CR terms, in Fig.~\ref{Fig8}, we plot the probability $P_{|g,0\rangle}(t)$ at various values of the modulation frequency $\nu$ with the system initially prepared in state $|g,0\rangle$. We see that the probability $P_{|g,0\rangle}(t)$ experiences fast oscillations and deviates from $1$ significantly in the absence of the modulation. When the modulation is applied, the magnitude of the oscillations decreases gradually with the increase of $\nu$. For sufficiently large $\nu$, the system almost remains in the state $|g,0\rangle$. These features can be understood by analyzing the interactions in the Hamiltonian $\tilde{H}(t)$. When $\nu=0$, i.e., without the modulation, the system will transit from the state $|g,0\rangle$ to other states with higher number of excitations due to the CR interactions. When the CR interactions are completely suppressed by the modulation, the system is well described by the JC Hamiltonian $\tilde{H}_{\textrm{JC}}(t)$. As a result, the system will stay in the ground state $|g,0\rangle$, which does not evolve under $\tilde{H}_{\textrm{JC}}(t)$.

\begin{figure}[tbp]
\center
\includegraphics[bb=13 3 282 353, width=0.47 \textwidth]{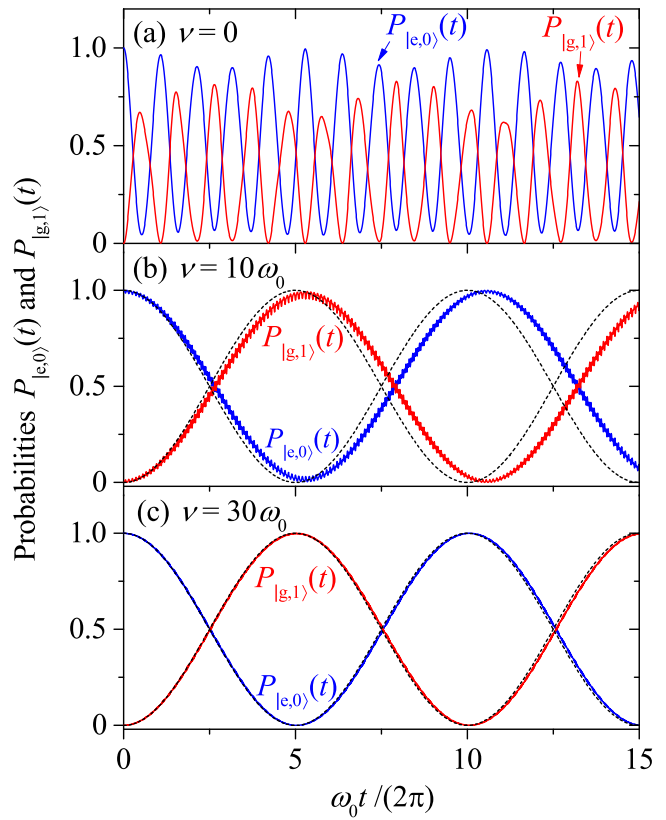}
\caption{(Color online) Time dependance of the probabilities $P_{|e,0\rangle}(t)$ (blue solid curves) and $P_{|g,1\rangle}(t)$ (red solid
curves) obtained with the exact Hamiltonian~(\ref{eq:H1}) at several values of $\nu$: (a) $\nu=0$, (b) $\nu=10\omega_{0}$, and (c) $\nu=30\omega_{0}$. The black short-dashed curves in panels (b) and (c) are the Rabi oscillations between the states $|e,0\rangle$ and $|g,1\rangle$ governed by the effective JC Hamiltonian~($\ref{eq:Vjc}$). Other parameters are: $g=0.5\omega_{0}$, $\xi=2.21868$ [$J_{0}(\xi)=0.1$], and $\omega_{c}=\omega_{0}$.}
\label{Fig9}
\end{figure}
We also consider the dynamics of this system starting from another initial state $|e,0\rangle$. In Fig.~\ref{Fig9}, we plot the probabilities $P_{|e,0\rangle}(t)$ and $P_{|g,1\rangle}(t)$ of the states $|e,0\rangle$ and $|g,1\rangle$, respectively, when the modulation frequency $\nu$ takes several values. The solid and short-dashed curves are obtained using the exact Hamiltonian~(\ref{eq:H1}) and the effective Hamiltonian~(\ref{eq:Vjc}), respectively. Figure~\ref{Fig9}(a) gives the state probabilities of the system in the absence of the modulation. In this case, Hamiltonian~(\ref{eq:H1}) reduces to a standard Rabi model in the interaction picture, and the probabilities of states $|e,0\rangle$ and $|g,1\rangle$ experience fast oscillations. More importantly, the total probability $(P_{|e,0\rangle}+P_{|g,1\rangle})$ of the system in the single-excitation subspace spanned by the basis states $|e,0\rangle$ and $|g,1\rangle$ is not normalized to unity due to the transitions induced by the CR terms. However, by introducing a properly-designed modulation under the condition~(\ref{RWAcod2}), the system can be well characterized by the effective JC Hamiltonian~($\ref{eq:Vjc}$). The probabilities $(P_{|e,0\rangle}$ and $P_{|g,1\rangle})$ then become a Rabi oscillation (independent of $\nu$) between the states $|e,0\rangle$ and $|g,1\rangle$, as shown by the black short-dashed curves in Figs.~\ref{Fig9}(b) and ~\ref{Fig9}(c). In Fig.~\ref{Fig9}(b), the small deviation of the approximate probabilities from the exact probabilities (solid curves) reflects the validity of the RWA. However, when the frequency $\nu$ is large enough, for example $\nu/\omega_{0}=30$ [Fig.~\ref{Fig9}(c)], the exact and approximate probabilities highly overlap with each other. In addition, the periods of the oscillations in Figs.~\ref{Fig9}(b) and~\ref{Fig9}(c) are longer than that in Fig.~\ref{Fig9}(a) due to the modification of the effective Rabi frequency by $J_{0}(\xi)$.

\section{\label{conclusion}Discussions and conclusions}

Finally, we present brief discussions on the implementation of this scheme with circuit-QED systems. The Rabi model can be realized by coupling a superconducting qubit with a transmission line resonator. Here the two-level system can be either a transmon qubit or a flux qubit. The resonance frequencies of the qubit and the resonator can be in the range of $5$ - $10$ GHz. The coupling strength between the qubit and resonator can be in either the strong- or the ultrastrong-coupling regimes (from hundreds of megahertz to some exceed one gigahertz). For example, $g/\omega_{c}\sim$ $0.05$ - $0.5$~\cite{Blais2009PRA,Gross2010NatPhy,Mooij2010PRL} in accordance with the parameters used in our scheme. The frequency modulation in the qubit Hamiltonian can be implemented by applying a periodic driving field on the qubit. Depending on the design of the qubit, the driving field can be introduced by using proper gate voltage or biased magnetic flux~\cite{Mooij2009,Porras2014PRL}. By tuning the driving amplitude and frequency, the modulation parameters $\xi$ and $\nu$ can be chosen on demand. Similar modulation schemes in experiments have been studied in~\cite{Porras2012PRL,Porras2014PRL}, and can be efficiently accessed following~\cite{Mooij2009}.

In conclusion, we have proposed a method to control the CR interactions in the quantum Rabi model by introducing a sinusoidal modulation to the frequency of the quantum two-level system. This control scheme includes the enhancement of the CR interactions in the JC regime and the suppression of these terms in the ultrastrong-coupling regime. By designing proper modulation frequency and amplitude, the rotating and CR interactions in the Rabi model
can be tailored to be either resonant or far off-resonant. In these cases, we have derived effective Hamiltonians to describe the dynamics of this system and verified detailed parameter conditions under which the approximations are valid. We have also studied the evolution of the state population in this system, which shows clear evidence of the manipulation of the interaction terms. In addition, we have investigated photon emission in the enhancement case by calculating the output photon flux rate. A continuous photon emission from the cavity vacuum in the JC regime is obtained.

\begin{acknowledgments}
J.F.H. is supported by the National Natural Science Foundation of China under Grants No. 11447102 and No. 11505055. J.Q.L. and L.T. are supported by the National Science Foundation under Award No. NSF-DMR-0956064. L.M.K. is supported by the National Fundamental Research Program of China (the 973 Program) under Grant No. 2013CB921804 and the National Natural Science Foundation of China under Grants No. 11375060, and No. 11434011.
\end{acknowledgments}


\begin{thebibliography}{10}
\bibitem{JCregimenote}    The weak (strong) coupling corresponds to the case in which the atom-field coupling strength is smaller (larger) than the decay rates of the atom and the field. In these two cases, the coupling strength should be much smaller than the frequencies of the atom and the field to confirm the validity of the RWA.
\end{thebibliography}
\end{document}